\documentstyle[preprint,aps,epsfig]{revtex}

\begin{document}
\draft
\title {Effective nonlinear optical properties of composite media of graded spherical particles}
\date {\today}
\author {L. Gao$^{1,2}$, J. P. Huang$^1$, and K. W. Yu$^1$}
\address { $^1$Department of Physics, The Chinese University of Hong Kong,\\ Shatin, New Territories, Hong Kong \\
  $^2$Department of Physics, Suzhou University, Suzhou 215 006, China}
\maketitle

\begin{abstract}

We have developed a nonlinear differential effective dipole
approximation (NDEDA), in an attempt to investigate the effective
linear and third-order nonlinear susceptibility of composite media
in which graded spherical inclusions with weak nonlinearity are
randomly embedded in a linear host medium. Alternatively, based on
a first-principles approach, we derived exactly the linear local
field inside the graded particles having power-law dielectric
gradation profiles. As a result, we obtain also the effective
linear dielectric constant and third-order nonlinear
susceptibility. Excellent agreement between the two methods is
numerically demonstrated. As an application, we apply the NDEDA to
investigate the surface plasma resonant effect on the optical
absorption, optical nonlinearity enhancement, and figure of merit
of metal-dielectric composites. It is found that the presence of
gradation in metal particles yields a broad resonant band in the
optical region, and further enhances the figure of merit.

{\bf PACS number(s):} 77.22.Ej, 42.65.-k, 42.79.Ry, 77.84.Lf





\end{abstract}


\newpage

\section{Introduction}

Graded materials, whose material properties can vary continuously
in space, are abundant in nature.  These materials have attracted
much interest as one of the advanced inhomogeneous composite
materials in various engineering applications~\cite{Yamanouchi}.
With the advent of fabrication techniques, these materials can be
well produced to tailor their properties for specific needs via
the design of the material and microstructure gradients. Such a
design makes graded materials quite different in physical
properties from the homogeneous materials and other conventional
composite materials. Moreover, the composite media consisting of
graded inclusions can be more useful and interesting than those of
homogeneous inclusions. Although various theories have been
established to investigate the optical and dielectric properties
of the composite media of homogeneous
inclusions~\cite{Jackson,AIP}, they fail to deal with the
inhomogeneous composites of graded inclusions. Recently, a
first-principles approach~\cite{Dong,Gu-JAP} and a differential
effective dipole approximation~\cite{Yu1,Huang} have been
presented in order to investigate the dielectric response of
graded materials.

The problem becomes more complicated by the presence of
nonlinearity in realistic composites. Besides inhomogeneity, such
nonlinearity plays also an important role in the effective
material properties of composite
media~\cite{Bergman,Shalaev,Hui,Sarychev}. It is thus necessary to
establish a new theory to study the effective nonlinear properties
of  graded composite media. In fact, the introduction of
dielectric gradation profiles in nonlinear composites is able to
provide an alternative way to control the local field fluctuation,
and hence let us obtain the desired effective nonlinear response.

In fact, the previous one-shell model~\cite{Gao1} and multi-shell
model~\cite{Haus}, which were used to study the effective
nonlinear optical property, can be seen as an initial model of
graded inclusions. In this paper, we will put forth a nonlinear
differential effective dipole approximation (NDEDA) to investigate
the effective linear and nonlinear dielectric properties of
composite media containing nonlinear graded spherical particles
(inclusions). For such particles, the linear and nonlinear
physical properties will continuously vary along their radius.

The paper is organized as follows. In section II, we describe the
model and define briefly the effective linear dielectric constant
and third-order nonlinear susceptibility. In section III, the
NDEDA is presented to investigate the effective linear dielectric
constant and third-order nonlinear susceptibility of nonlinear
graded composite media in the dilute limit. In section IV, based
on a first-principles approach, we derive the exact solutions for
composite media having power-law gradation profiles inside the
inclusions, which is followed by the numerical results in section
V. Finally, some conclusion and discussion is shown in Section VI.

\section{Model and definition of effective linear and nonlinear responses}

Let us consider a nonlinear composite system, in which identical
graded spherical inclusions with radius $a$, are randomly embedded
in a linear host medium of dielectric constant $\epsilon_2$. The
local constitutive relation between the displacement $({\bf D})$
and the electric field $({\bf E})$ inside the graded particle is
given by
\begin{equation}
{\bf D}=\epsilon(r){\bf E}+\chi(r)|{\bf E}|^2{\bf
E},\label{relation}
\end{equation}
where $\epsilon(r)$ and $\chi(r)$ are, respectively, the linear
dielectric constant and third-order nonlinear susceptibility. Note
both $\epsilon(r)$ and $\chi(r)$ are radial functions. Here we
assume that the weak nonlinearity condition is
satisfied~\cite{Stroud}. In other words, the contribution of the
second (nonlinear) part [$\chi_s( r)|{\bf E}|^2$] in the
right-hand side of Eq.~(\ref{relation}) is much less than that of
the first (linear) part $\epsilon(r)$. We restrict further our
discussion to the quasi-static approximation, under which the
whole composite medium can be regarded as an effective homogeneous
one with effective linear dielectric constant $\epsilon_e$ and
effective third-order nonlinear susceptibility $\chi_e$. To show
the definitions of  $\epsilon_e$ and $\chi_e$, we
have~\cite{Stroud}
\begin{equation}
\langle {\bf D}\rangle =\epsilon_e {\bf E}_0 +\chi_e|{\bf
E}_0|^2{\bf E}_0,
\end{equation}
where $\langle\cdot \cdot \cdot\rangle$ represents the spatial
average, and ${\bf E}_0=E_0{\bf e}_z$ is the external applied
field along $z$ axis.

The effective linear dielectric constant $\epsilon_e$ is given by
\begin{equation}
\epsilon_e{\bf E}_0=\frac{1}{V}\int_V \epsilon_i{\bf E}_{{\rm
lin,i}}{\rm d}V= f\langle\epsilon(r){\bf E}_{{\rm lin,1}}\rangle
+(1-f)\epsilon_2\langle{\bf E}_{{\rm lin,2}}\rangle,\label{Ee1}
\end{equation}
where $f$ is the volume fraction of the graded particles and the
subscript stands for the linear local field [i.e., obtained for
the same system but with $\chi(r)=0$].

In view of the existence of nonlinearity inside the graded
particles, $\chi_e$ can then be written as~\cite{Stroud,Wood}
\begin{equation}
\chi_e{\bf E}_0^4=\frac{1}{V}\int_V \chi_i|{\bf E}|_{{\rm
lin,i}}^2{\bf E}_{{\rm lin,i}}^2 {\rm d}V=\frac{1}{V}
\int_{\Omega_i}\chi(r)|{\bf E}|^2_{{\rm lin,1}}{\bf E}^2_{{\rm
lin,1}}{\rm d}V= f\langle\chi( r)|{\bf E}|^2_{{\rm lin,1}}{\bf
E}^2_{{\rm lin,1}}\rangle.\label{definitionXe}
\end{equation}

In the next section, we will develop a NDEDA (nonlinear
differential effective dipole approximation), in an attempt to
derive the equivalent linear dielectric constant
$\bar{\epsilon}(a)$ and third-order nonlinear susceptibility
$\bar{\chi}(a)$ of the nonlinear graded inclusions. Then, the
effective linear dielectric constant and third-order nonlinear
susceptibility of the composite media of nonlinear graded
inclusions will be derived accordingly in the dilute limit.

\section{Nonlinear differential effective dipole approximation}

To establish the NDEDA, we first mimic the gradation profile by a
multi-shell constriction. That is, we build up the dielectric
profile by adding shells gradually~\cite{Yu1}.  We start with an
infinitesimal spherical core with linear dielectric constant
$\epsilon(0)$ and third-order nonlinear susceptibility $\chi(0)$,
and keep on adding spherical shells with linear dielectric
constant $\epsilon(r)$ and third-order nonlinear susceptibility
$\chi(r)$ at radius $r$, until $r=a$ is reached. At radius $r$,
the inhomogeneous spherical particle with space-dependent
dielectric gradation profiles $\epsilon(r)$ and $\chi(r)$ can be
replaced by a {\it homogenous} sphere with the equivalent
dielectric properties $\bar{\epsilon}(r)$ and $\bar{\chi}(r)$.
Here the {\it homogeneous} sphere should induce the same dipole
moment as the original inhomogeneous sphere.

Next, we add to the sphere a spherical shell of infinitesimal
thickness ${\rm d}r$, with dielectric constant $\epsilon(r)$ and
nonlinear susceptibility $\chi(r)$. In this sense, the coated
inclusions is composed of a spherical core with radius $r$, linear
dielectric constant $\bar{\epsilon}(r)$ and nonlinear
susceptibility $\bar{\chi}(r)$, and a shell with outermost radius
$r+{\rm d}r$, linear dielectric constant $\epsilon(r)$ and
nonlinear susceptibility $\chi(r)$. Since these coated inclusions
are randomly embedded in a linear host medium, under the
quasi-static approximation, we can readily obtain the linear
electric potentials in the core, shell and host medium by solving
the Laplace equation~\cite{Scaife}
\begin{eqnarray}
\phi_c&=&-E_0AR\cos\theta, \qquad R<r, \nonumber \\
\phi_s&=&-E_0\left(BR-\frac{Cr^3}{R^2}\right)\cos\theta, \qquad r<R<r+{\rm d}r, \\
\phi_h&=&-E_0\left(R-\frac{D(r+{\rm d}r)^3}{R^2}\right)\cos\theta,
\qquad R>r+{\rm d} r, \nonumber
\end{eqnarray}
where
\begin{eqnarray}
A&=&\frac{9\epsilon_2\epsilon(r)}{Q}, \qquad
B=\frac{3\epsilon_2[\bar{\epsilon}(r)+2\epsilon(r)]}{Q}, \quad
\quad
  C=\frac{3\epsilon_2[\bar{\epsilon}(r)-\epsilon(r)]}{Q}, \nonumber \\
D&=&\frac{[\epsilon(r)-\epsilon_2][\bar{\epsilon}(r)+2\epsilon(r)]+
\lambda[\epsilon_2+2\epsilon(r)][\bar{\epsilon}(r)-\epsilon(r)]}{Q},\nonumber
\end{eqnarray}
with interfacial parameter $\lambda\equiv [r/(r+{\rm d}r)]^3$, and
$$
Q=[\epsilon(r)+2\epsilon_2][\bar{\epsilon}(r)+2\epsilon(r)]+
2\lambda[\epsilon(r)-\epsilon_2][\bar{\epsilon}(r)-\epsilon(r)].
$$

The effective (overall) linear dielectric constant of  the system
is determined by the dilute-limit expression~\cite{Yu2}
\begin{equation}
\epsilon_e=\epsilon_2+3p\epsilon_2D,
\end{equation}
where $p$ is the volume fraction of graded particles with radius
$r$.  The equivalent dielectric constant $\bar{\epsilon}(r+dr)$
for the graded particles with radius $r+{\rm d}r$ can be obtained
self-consistently by the vanishing of the dipole factor $D$ by
replacing $\epsilon_2$ with $\bar{\epsilon}(r+{\rm d}r)$. Taking
the limit ${\rm d}r\rightarrow 0$ and keeping to the first order
in ${\rm d}r$, we obtain
 \begin{eqnarray}
 \bar{\epsilon}(r+dr)&=&
 \epsilon(r)+3\epsilon(r)\lambda\cdot\frac{\bar{\epsilon}(r)-\epsilon(r)}
 {\bar{\epsilon}(r)(1-\lambda)+\epsilon(r)(2+\lambda)} \nonumber \\
 &=&\bar{\epsilon}(r)-\frac{\bar{\epsilon}(r)-\epsilon(r)}{r}\cdot\left[3+
 \frac{\bar{\epsilon}(r)-\epsilon(r)}{\epsilon(r)}\right]{\rm d}r.
 \end{eqnarray}
 Thus, we have the differential equation for the equivalent dielectric
 constant $\bar{\epsilon}(r)$ as~\cite{Yu1}
 \begin{equation}
 \frac{{\rm d}\bar{\epsilon}(r)}{{\rm d}r}=
 \frac{[\epsilon(r)-\bar{\epsilon}(r)]\cdot[\bar{\epsilon}(r)+2\epsilon(r)]}
 {r\epsilon(r)}.\label{Tartar1}
 \end{equation}
 Note that Eq.~(\ref{Tartar1}) is just the Tartar formula, derived for assemblages of spheres with varying
 radial and tangential conductivity~\cite{Milton}.

 Next, we speculate on how to derive the equivalent nonlinear susceptibility $\bar{\chi}(r)$.
 After applying Eq.~(\ref{definitionXe}) to the coated particles with radius $r+{\rm d}r$, we have
\begin{equation}
 \bar{\chi}(r+{\rm d}r)\frac{\langle|{\bf E}|^2{\bf E}^2\rangle_{R\le r+{\rm d}r}}{|{\bf E}_0|^2{\bf E}_0^2}
 =\lambda
 \bar{\chi}(r)\frac{\langle|{\bf E}|^2{\bf E}^2\rangle_{R\le r}}{|{\bf E}_0|^2{\bf E}_0^2}
 +(1-\lambda)
 \frac{\langle\chi(r)|{\bf E}|^2{\bf E}^2\rangle_{r< R\le r+{\rm d}r}}{|{\bf E}_0|^2{\bf
 E}_0^2}.\label{coatedXe1}
 \end{equation}
As ${\rm d}r\rightarrow 0$, the left-hand side of the above
equation admits
\begin{eqnarray}
\bar{\chi}(r+{\rm d}r)\frac{\langle\left|{\bf E}\right|^2{\bf
E}^2\rangle_{R \le r+{\rm d}r}}{|{\bf E}_0|^2{\bf E}_0^2}
&=&\bar{\chi}(r+{\rm d}r) \left|\frac{3\epsilon_2}
{\bar{\epsilon}(r+{\rm d}r)+2\epsilon_2}\right|^2
\left(\frac{3\epsilon_2}
{\bar{\epsilon}(r+{\rm d}r)+2\epsilon_2}\right)^2 \nonumber \\
&=&\bar{\chi}(r)|K|^2K^2-{\rm d}r\bar{\chi}(r)|K|^2K^2 \left[
\frac{3{\rm d}\bar{\epsilon}(r)/{\rm d}r}
{2\epsilon_2+\bar{\epsilon}(r)}+
\left(\frac{{\rm d}\bar{\epsilon}(r)/{\rm d}r}{2\epsilon_2+\bar{\epsilon}(r)}\right)^*\right] \nonumber \\
&&+|K|^2K^2\frac{{\rm d} \bar{\chi}(r)}{{\rm d}r}\cdot{\rm
d}r,\label{left1}
\end{eqnarray}
with $K=(3\epsilon_2)/[ \bar{\epsilon}(r)+2\epsilon_2].$ The first
part of the right-hand side of Eq.~(\ref{coatedXe1}) is written as
\begin{equation}
\lambda
 \frac{\bar{\chi}(r)\langle|{\bf E}|^2{\bf E}^2\rangle_{R\le r}}{|{\bf E}_0|^2{\bf E}_0^2}=
\bar{\chi}(r)|K|^2K^2\left[1+(6y+2y^*-3)\frac{{\rm
d}r}{r}\right],\label{right1}
\end{equation}
where
$$
y=\frac{[\epsilon(r)-\epsilon_2]
[\bar{\epsilon}(r)-\epsilon(r)]}{\epsilon(r)[\bar{\epsilon}(r)+2\epsilon_2]}.
$$
The second part of the right-hand side of Eq.~(\ref{coatedXe1})
has the form~\cite{Yu2}
\begin{eqnarray}
(1-\lambda) \frac{\langle\chi(r)|{\bf E}|^2{\bf E}^2\rangle_{r< R
\le r+{\rm d}r}}{|{\bf E}_0|^2{\bf E}_0^2} &=&
\frac{3\chi(r)}{5r}{\rm d}r|z|^2z^2 \nonumber\\&
&\times(5+18x^2+18|x|^2+4x^3+12x|x|^2+
 24|x|^2x^2),\label{right2}
 \end{eqnarray}
 where
 $$
 x=\frac{\bar{\epsilon}(r)-\epsilon(r)}{\bar{\epsilon}(r)+2\epsilon(r)} \quad {\rm and} \quad
 z=\frac{\epsilon_2[\bar{\epsilon}(r)+2\epsilon(r)]}{\epsilon(r)[\bar{\epsilon}(r)+2\epsilon_2]}.
$$

Substituting Eqs.~(\ref{left1}),~(\ref{right1})~and~(\ref{right2})
into Eq.~(\ref{coatedXe1}), we have a differential equation for
the equivalent nonlinear susceptibility $\bar{\chi}(r)$, namely,
\begin{eqnarray}
\frac{{\rm d}\bar{\chi}(r)}{{\rm
d}r}&=&\bar{\chi}(r)\left[\frac{3{\rm d}\bar{\epsilon}(r)/{\rm
d}r}{2\epsilon_2+ \bar{\epsilon}(r)} +\left( \frac{{\rm
d}\bar{\epsilon}(r)/{\rm
d}r}{2\epsilon_2+\bar{\epsilon}(r)}\right)^*\right]+
\bar{\chi}(r)\cdot\frac{6y+2y^*-3}{r}
+\frac{3\chi(r)}{5r}\cdot\left|\frac{\bar{\epsilon}(r)+2\epsilon(r)}{3\epsilon(r)}\right|^2
\nonumber \\
&&\cdot
\left(\frac{\bar{\epsilon}(r)+2\epsilon(r)}{3\epsilon(r)}\right)^2(5+18x^2+18|x|^2+4x^3+12x|x|^2+24|x|^2x^2).\label{NDEDA1}
\end{eqnarray}

So far, the equivalent $\bar{\epsilon}(r)$ and $\bar{\chi}(r)$ of
graded spherical particles of radius $r$ can be calculated, at
least numerically,  by solving the differential equations
Eqs.~(\ref{Tartar1})~and~(\ref{NDEDA1}), as long as $\epsilon(r)$
(dielectric-constant gradation profile) and  $\chi(r)$
(nonlinear-susceptibility gradation profile) are given. Here we
would like to mention that, even though $\chi(r)$ is independent
of $r$, the equivalent $\bar{\chi}(r)$ should still be dependent
on $r$ because of $\epsilon(r)$ as a function of $r$. Moreover,
for both $\epsilon(r)=\epsilon_1$ and $\chi(r)=\chi_1$ (i.e., they
are both constant and independent of $r$),
Eqs.~(\ref{Tartar1})~and~(\ref{NDEDA1}) will naturally reduce to
the solutions $\bar{\epsilon}(r)=\epsilon_1$ and
$\bar{\chi}(r)=\chi_1$.

To obtain $\bar{\epsilon}(r=a)$ and $\bar{\chi}(r=a)$, we
integrate Eqs.~(\ref{Tartar1})~and~(\ref{NDEDA1}) numerically at
given initial conditions $\bar{\epsilon}(r\rightarrow 0)$ and
$\bar{\chi}(r\rightarrow 0)$. Once $\bar{\epsilon}(r=a)$ and
$\bar{\chi}(r=a)$ are calculated, we can take one step forward to
work out the effective linear and nonlinear responses $\epsilon_e$
and $\chi_e$ of the whole composite in the dilute limit,
i.e.~\cite{Stroud},
\begin{equation}
\epsilon_e=\epsilon_2+3\epsilon_2f\frac{\bar{\epsilon}(r=a)-\epsilon_2}{\bar{\epsilon}(r=a)+2\epsilon_2},
\end{equation}
and
\begin{equation}
\chi_e=f\bar{\chi}(r=a)\left|\frac{3\epsilon_2}{\bar{\epsilon}(r=a)+2\epsilon_2}
\right|^2\left(\frac{3\epsilon_2}
{\bar{\epsilon}(r=a)+2\epsilon_2}\right)^2.
\end{equation}

\section{Exact solution for power-law gradation profiles}

Based on the first-principles approach, we have found that, for a
power-law dielectric gradation profile, i.e.,
$\epsilon(r)=A(r/a)^n$, the potential in the graded inclusions and
the host medium can be exactly given by~\cite{Dong}
\begin{eqnarray}
\phi_i(r) &=& -\xi_1E_0r^s\cos\theta, \quad r<a, \nonumber \\
\phi_h(r) &=& -E_0r\cos\theta+\frac{\xi_2}{r^2}E_0\cos\theta,
\quad r>a,
\end{eqnarray}
where the coefficients $\xi_1$ and $\xi_2$ have the form
$$
\xi_1=\frac{3a^{1-s}\epsilon_2}{sA+2\epsilon_2}
\quad\mathrm{and}\quad
\xi_2=\frac{sA-\epsilon_2}{sA+2\epsilon_2}a^3,
$$
and $s$ is given by
$$
s=\frac{1}{2}\left[\sqrt{9+2n+n^2} -(1+n)\right].
$$

The local electric field inside the graded inclusions can be
derived from the potential ${\bf E}=-\nabla\phi$,
\begin{eqnarray}
{\bf E}_i&=&\xi_1E_0r^{s-1}(s\cos\theta{\bf e}_r-\sin\theta{\bf
e}_\theta) =\xi_1E_0r^{s-1}\{ (s-1)\cos\theta\sin\theta\cos\phi
{\bf e}_x \nonumber \\
&&+(s-1)\cos\theta\sin\theta\sin\phi{\bf e}_y
+[(s-1)\cos^2\theta+1]{\bf e}_z\},\label{local1}
\end{eqnarray}
where ${\bf e}_r$, ${\bf e}_\theta$, ${\bf e}_x$, ${\bf e}_y$ and
${\bf e}_z$ are unix vectors. In the dilute limit, from
Eq.~(\ref{Ee1}), we can obtain the effective linear dielectric
constant as follows
\begin{eqnarray}
\epsilon_e&=&\epsilon_2+\frac{1}{VE_0}\int_{\Omega_i}\left[A(r/a)^n-\epsilon_2\right]{\bf e}_z\cdot{\bf E}_i{\rm d}V \nonumber \\
       &=&\epsilon_2+3\epsilon_2f\frac{2+s}{sA+2\epsilon_2}
       \left(\frac{A}{2+n+s}-\frac{\epsilon_2}{2+s}\right).\label{exact1}
\end{eqnarray}
On the other hand, the substitution of Eq.~(\ref{local1}) into
Eq.~(\ref{definitionXe}) yields
\begin{eqnarray}
\chi_e&=&\frac{1}{V}\int_{\Omega_i}\chi(r)|\xi_1|^2\xi_1^2(s^2\cos^2\theta+\sin^2\theta)^2r^{4s-2}\sin\theta
{\rm d}r{\rm d}\theta{\rm d}\phi
\nonumber \\
&=&\frac{f}{5a^3}|\xi_1|^2\xi_1^2(8+4s+3s^4)\cdot\int_0^a\chi(r)r^{4s-2}{\rm
d}r.\label{ChiE1}
\end{eqnarray}
For example, for a linear profile of $\chi(r)$, i.e.,
$\chi(r)=k_1+k_2\cdot r/a$, Eq.~(\ref{ChiE1}) leads to
\begin{equation}
\chi_e=\frac{f}{20}\left|\frac{3\epsilon_2}{sA+2\epsilon_2}\right|^2
\left(\frac{3\epsilon_2}{sA+2\epsilon_2}\right)^2(8+4s^2+3s^4)
\left(\frac{k_2}{s}+\frac{4k_1}{4s-1}\right).\label{exactEe1}
\end{equation}
In addition, for a power-law profile of $\chi(r)$, namely,
$\chi(r)=k_1(r/a)^{k_2}$, Eq.~(\ref{ChiE1}) produces
\begin{equation}
\chi_e=\frac{f}{5}\left|\frac{3\epsilon_2}{sA+2\epsilon_2}\right|^2
\left(\frac{3\epsilon_2}{sA+2\epsilon_2}\right)^2
 k_1\left(\frac{8+4s^2+3s^4}{k_2-1+4s}\right).\label{exactXe1}
\end{equation}

\section{Numerical results}

We are now in a position to evaluate the NDEDA.  For the
comparison between the first-principles approach and the NDEDA, we
first perform numerical calculations for the case where the
dielectric constant exhibits power-law gradation profiles
$\epsilon(r)=A(r/a)^n$, while the third-order nonlinear
susceptibility shows two model gradation profiles: (a) linear
profile $\chi(r)=k_1+k_2\cdot r/a$, and (b) power-law profile
$\chi(r)=k_1(r/a)^{k_2}$. Without loss of generality, we take
$\epsilon_2=1$ and $a=1$ for numerical calculations. The
fourth-order Runge-Kutta algorithm is adopted to integrate the
differential equations [Eqs.~(\ref{Tartar1})~and~(\ref{NDEDA1})]
with step size $0.01$. Meanwhile, the initial core radius is set
to be $0.001$. It was verified that this step size guarantees
accurate numerics.

In Fig.~1, the effective linear dielectric constant ($\epsilon_e$)
is plotted as a function of $A$ for various indices $n$. It is
shown that $\epsilon_e$ exhibits a monotonic increase for
increasing $A$ (and decreasing $n$). This can be understood by
using the equivalent dielectric constant $\bar{\epsilon}(r=a)$
which increases as $A$ increases ($n$ decreases). Moreover, the
excellent agreement between the NDEDA [Eq.~(\ref{Tartar1})] and
the first-principles approach [Eq.~(\ref{exact1})] is shown as
well.

Next, the effective third-order nonlinear susceptibility
($\chi_e$) is plotted as a function of $A$ for the linear
gradation profile $\chi(r)=k_1+k_2\cdot r/a$ (Fig.~2), and for the
power-law profile $\chi(r)=k_1(r/a)^{k_2}$ (Fig.~3). We find that
the effective nonlinear susceptibility decreases for increasing
$A$. The reason is that, as mentioned above,
 for larger $A$, the graded inclusions
possess larger equivalent dielectric constant, and the local field
inside the nonlinear inclusions will become more weak, which
results in a weaker effective nonlinear susceptibility ($\chi_e$).
In addition, increasing $n$ leads generally to increasing
$\chi_e$, and such a trend is clearly observed at large $A$.
Again, we obtain the excellent agreement between
 the first-principles approach [Eqs.~(\ref{exactEe1})~and~(\ref{exactXe1})]
 and the NDEDA [Eqs.~(\ref{Tartar1})~and~(\ref{NDEDA1})].

In what follows, we investigate the surface plasma resonance
effect on the metal-dielectric composite. We adopt the Drude-like
 dielectric constant for graded metal particles, namely,
\begin{equation}
\epsilon(r)=1-\frac{\omega^2_p(r)}{\omega[\omega+{\rm
i}\gamma(r)]},
\end{equation}
where $\omega_p(r)$ and $\gamma(r)$ are the radius-dependent
plasma frequency  and damping coefficient, respectively. For the
sake of simplicity, set $\chi(r)=\chi_1$ to be independent of $r$,
in an attempt to emphasize the enhancement of the effective
optical nonlinearity, and
  $\epsilon_2=1.77$ (the dielectric constant of water).
We assume further $\omega_p(r)$ to be
\begin{equation}
\omega_p(r)=\omega_p(1-k_{\omega}\cdot\frac{r}{a}), \quad r<a.
\end{equation}
This form is quite physical for $k_\omega>0$, since the center of
grains can be
 better metallic so that $\omega_p(r)$ is larger,
while the boundary of the grain may be poorer metallic so that
$\omega_p(r)$ is much smaller. Such the variation can also appear
because of the temperature effect~\cite{Gao2}. For small
particles, we have the radius-dependent $\gamma(r)$
as~\cite{Neeves}
\begin{equation}
\gamma(r)=\gamma(\infty)+\frac{k_\gamma}{r/a}, \quad r<a,
\end{equation}
where $\gamma(\infty)$ stands for the damping coefficient  in the
bulk material. Here $k_{\gamma}$ is a constant which is related to
the Fermi velocity $v_F$. In this case, the exact solution being
predicted by a first-principles approach is absent. Fortunately,
we can resort to the NDEDA instead.

In Fig.3, we plot the optical absorption [$\sim {\rm
Im}(\epsilon_e)$], the modulus of the effective third-order
optical nonlinearity enhancement ($|\chi_e|/\chi_1$) and the
figure of merit
 ($|\chi_e|/{\rm Im}(\epsilon_e)$) versus the incident angular frequency $\omega$.
For various variance slopes $k_\omega$, large surface plasma
resonance peaks can appear around $\omega\approx 0.4\omega_p,$ as
expected. However, when the radius dependence of $\omega_p(r)$ is
taken into account (i.e., $k_\omega\ne 0$), besides a sharp peak,
a broad continuous resonant bands in the high frequency region is
clearly observed. The broad spectra result from the effect of
radius dependence of the plasma frequency. In Ref.\cite{Gao1}, we
found that, when the shell model is considered, there should be a
broad continuous spectrum around the large pole in the spectral
density function. In fact, the graded particles under
consideration can be regarded as a certain limit of multi shells,
which hence should lead to broader spectra in ${\rm
Im}(\epsilon_e)$, $|\chi_e|/\chi_1$ as well as $|\chi_e|/{\rm
Im}(\epsilon_e)$. In addition, we note that, with increasing
$k_\omega$, both the surface plasma frequency and the center of
resonant bands are red-shifted. In particular, the resonant bands
can become more broad due to strong inhomogeneity of the
particles. From the figure, we conclude that, although the
third-order optical nonlinearity is always accompanied with the
optical absorption, the figure of merit in the high frequency
region is still attractive due to the presence of {\it weak}
optical absorption. Thus, we believe that graded particles may
have potential applications in obtaining the optimal figure of
merit, and make the composite media more realistic for practical
applications.

Finally, we focus on the effect of $\gamma(r)$ on the nonlinear
optical property in Fig.~5. As evident from the results, the
variation of $k_\gamma$ plays an important role in the magnitude
of the effective optical properties, particularly at the surface
plasma resonance frequency.

\section{Conclusion and discussion}

Here a few comments are in order. In this work, we have developed
an NDEDA (nonlinear differential effective dipole approximation)
to calculate the effective linear and nonlinear dielectric
responses of composite media containing nonlinear graded
inclusions. The results obtained from the NDEDA are compared with
the exact solutions derived from a first-principles approach for
the power-law dielectric gradation profiles, and the excellent
agreement between them has been shown. We should remark that the
exact solutions are also obtainable for the linear dielectric
gradation profiles with small slopes (the derivation not shown
here). In this case, the excellent agreement between the two
methods can be shown as well since the NDEDA is valid indeed for
arbitrary gradation profiles. In general, the exact solution is
quite few in realistic composite research, and thus our NDEDA can
be used as a benchmark.

It is instructive to develop the first-principles approach to
nonlinear graded composites. The perturbation
approach~\cite{Gu-Yu92} in weakly nonlinear composites as well as
the variational approach~\cite{Yu-Gu94} in strongly nonlinear
composites are just suitable for this problem.

The NDEDA is strictly valid in the dilute limit. To achieve the
strong optical nonlinearity enhancement, we need possibly
nonlinear inclusions with high volume fractions. In this
connection, the effect of the volume fraction is expected to cause
a further broadening of the resonant peak, and possibly, a desired
separation of the optical absorption peak from the nonlinearity
enhancement due to mutual interactions~\cite{Gao3}. Therefore, it
is of particular interest to generalize the NDEDA for treating the
case of high volume fractions.

To one's interest, the NDEDA can be applied to biological cells.
Since the interior of biological cells is often inhomogeneous and
nonlinear in nature, they can be treated as particles having
dielectric gradation profiles~\cite{Huang}. Moreover, in the case
of cell membranes containing mobile charges introduced by the
adsorbed hydrophobic ions, the local dielectric anisotropy occurs
naturally, and should be expected to play a
role~\cite{Zimmermann}. Thus, it is also interesting to see what
happens to the NDEDA as one takes into account the local
dielectric anisotropy. The resultant anisotropic NDEDA will help
to investigate the AC electrokinetic phenomena of biological
cells~\cite{Gao4}. Work along this direction is in progress.

To sum up, we put forth an NDEDA and a first-principles approach
for investigating the optical responses of nonlinear graded
spherical particles. The excellent agreement between the two
methods has been shown. As an application, we applied the NDEDA to
discuss the surface plasma resonance effect on the effective
linear and nonlinear optical properties like the optical
absorption, the optical nonlinearity enhancement, and the figure
of merit. It is found that the dielectric gradation profile can be
used to control the surface plasma resonance and achieve the large
figure of merit in the high-frequency region, where the optical
absorption is quite small.

\section*{Acknowledgment}

This work has been supported by the Research Grants Council of the
Hong Kong SAR Government under project numbers CUHK 4245/01P and
CUHK 403303, and by the National Natural Science Foundation of
China under Grant No.~10204017 (L.G.) and  the  Natural Science of
Jiangsu Province under Grant No.~BK2002038 (L.G.).

\begin{figure}[h]
\caption{ The effective linear dielectric constant $\epsilon_e$
versus $A$ for the power-law dielectric gradation profile
$\epsilon(r)=A(r/a)^n$ in the dilute limit $f=0.05$.  Lines:
numerical results from the NDEDA [Eq.~(\ref{Tartar1})]; Symbols:
exact results [Eq.~(\ref{exact1})].}
\end{figure}

\begin{figure}[h]
\caption{ The effective third-order nonlinear susceptibility
$\chi_e$ versus $A$ for power-law dielectric-constant gradation
profile $\epsilon(r)=A(r/a)^n$ and linear nonlinear-susceptibility
gradation profile $\chi(r)=k_1+k_2\cdot r/a$ with (a) $k_1=1$ and
$k_2=1$, and (b) $k_1=2$ and $k_2=3$. Lines: numerical results
from the NDEDA [Eqs.~(\ref{Tartar1})~and~(\ref{NDEDA1})]; Symbols:
exact results [Eq.~(\ref{ChiE1})].}
\end{figure}

\begin{figure}[h]
\caption{Same as Fig.2, but for power-law nonlinear-susceptibility
gradation profile $\chi(r)=k_1(r/a)^{k_2}$.}
\end{figure}

\begin{figure}[h]
\caption{ (a) The linear optical absorption ${\rm
Im}(\epsilon_e)$, (b) the enhancement of the third-order optical
nonlinearity $|\chi_e|/\chi_1$, and (c) the figure of merit
$\equiv |\chi_e|/{\rm Im}(\epsilon_e)$  versus the incident
angular frequency $\omega/\omega_p$ for dielectric-constant
gradation profile
$\epsilon(r)=1-\omega_p^2(r)/[\omega(\omega+i\gamma(r))]$ with
$\omega_p(r)=\omega_p(1-k_\omega\cdot r/a)$ and
$\gamma(r)=0.01\omega_p$. Parameters: $\epsilon_2=1.77$ and
$f=0.05$.}
\end{figure}

\begin{figure}[h]
\caption{Same as Fig.4, but with $\omega_p(r)=\omega_p$ and
$\gamma(r)=\gamma(\infty)+k_{\gamma}/(r/a)$ for
$\gamma(\infty)=0.01\omega_p$.}
\end{figure}

\newpage
\centerline{\epsfig{file=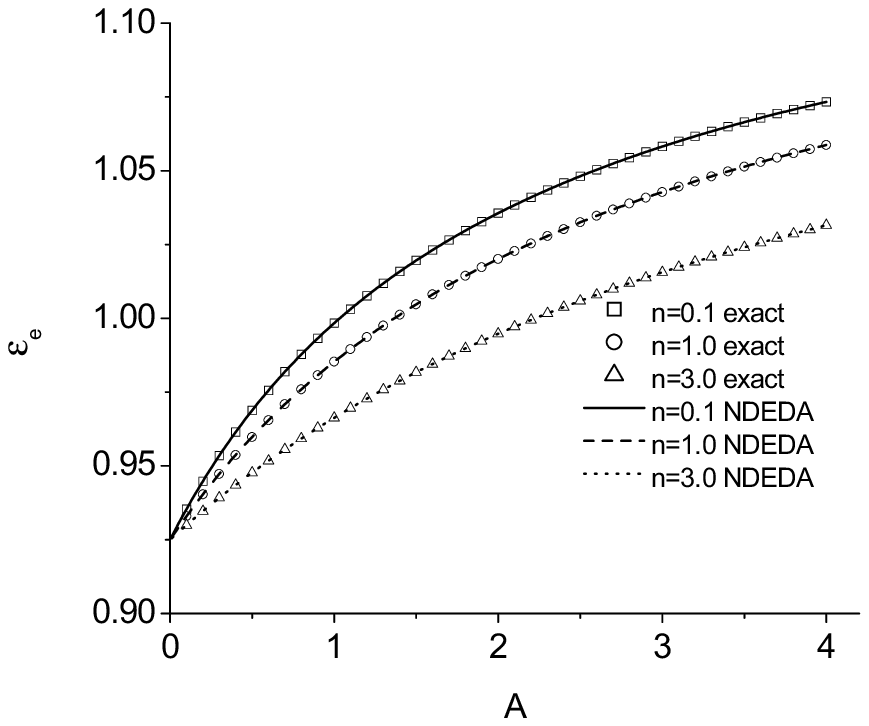,width=400pt}}
\centerline{Fig.1./Gao, Huang and Yu}

\newpage
\centerline{\epsfig{file=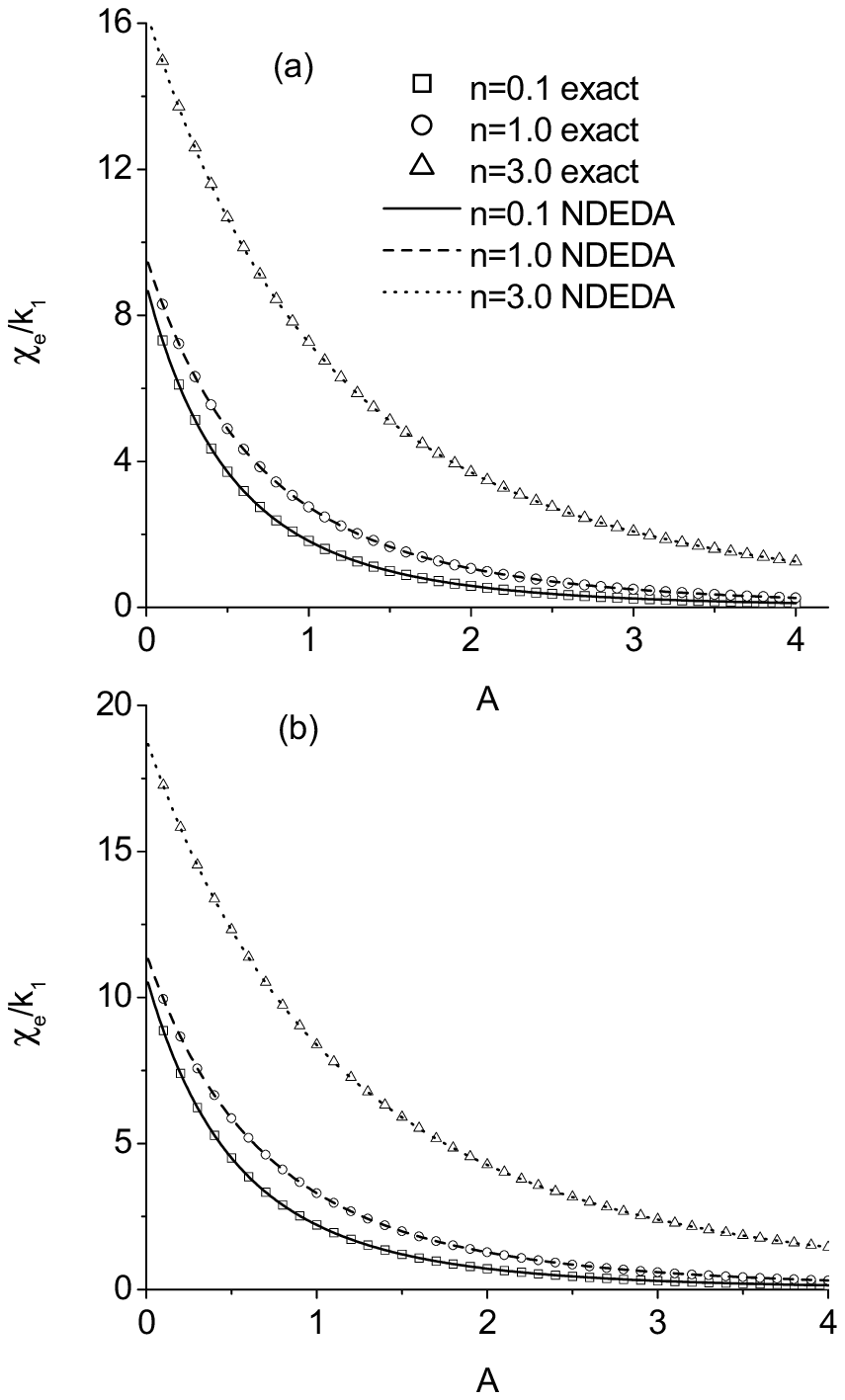,width=400pt}} \centerline{Fig.2.
/Gao, Huang and Yu}

\newpage
\centerline{\epsfig{file=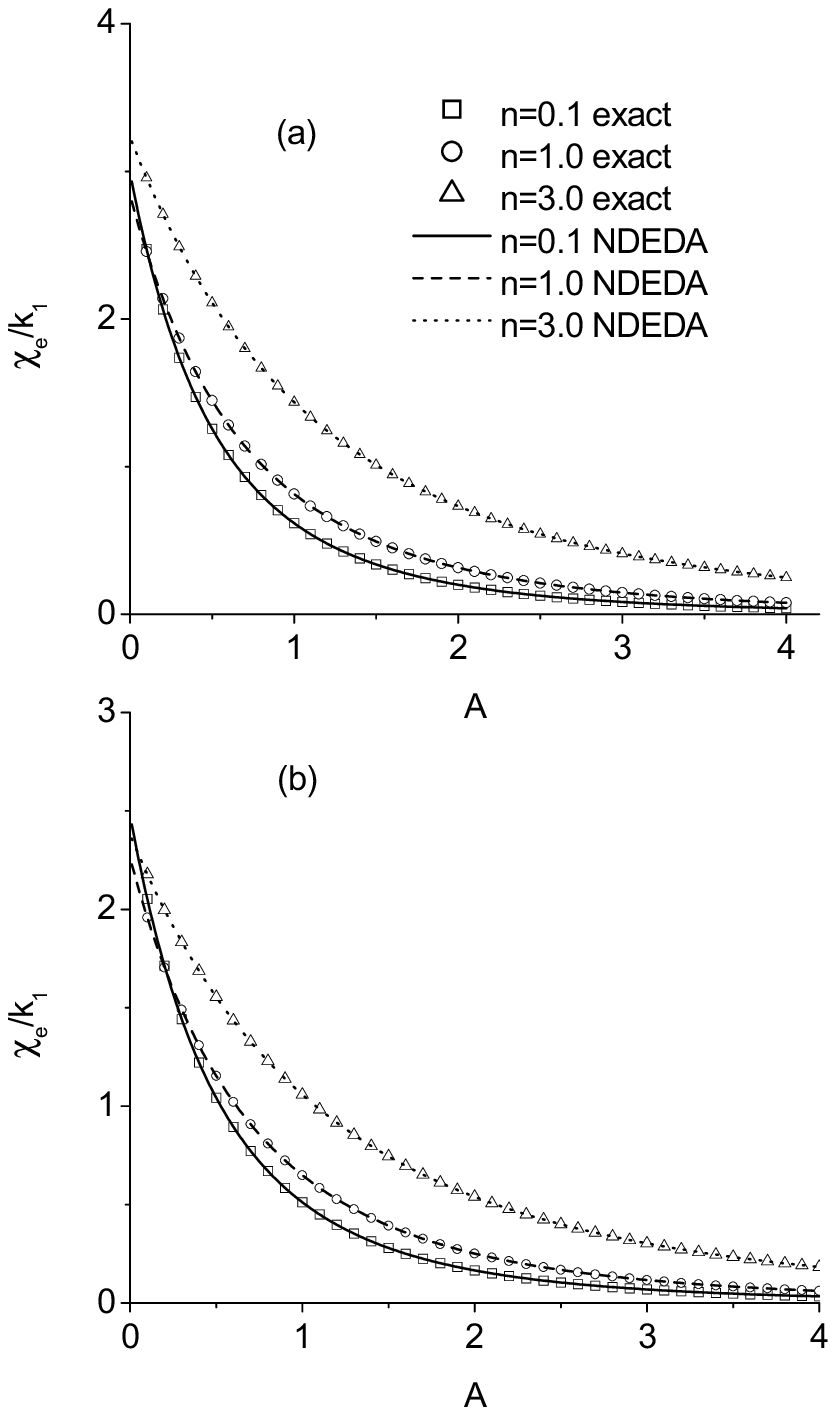,width=400pt}} \centerline{Fig.3.
/Gao,Huang and Yu}

\newpage
\centerline{\epsfig{file=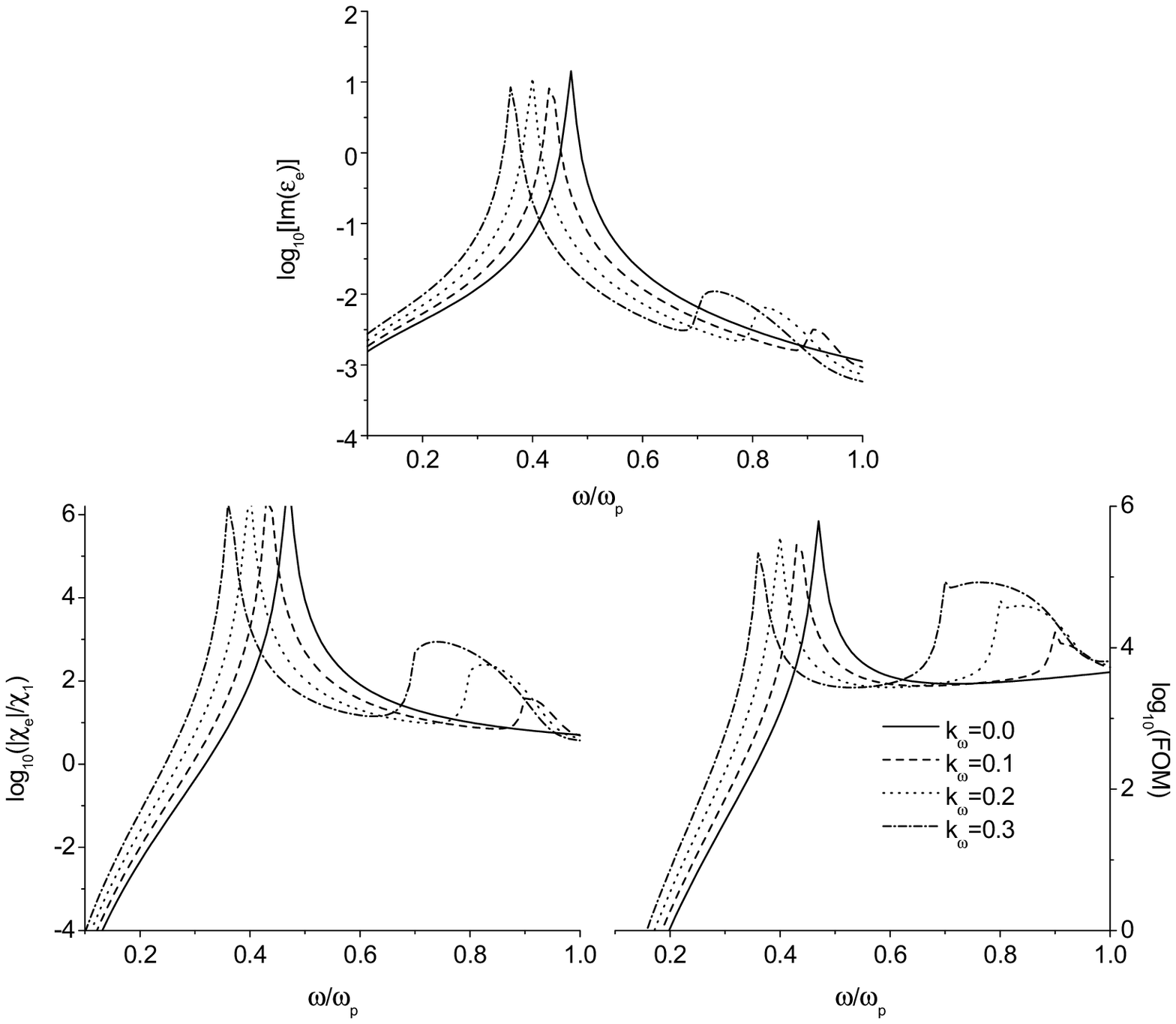,width=400pt}} \centerline{Fig.4.
/Gao, Huang and Yu}

\newpage
\centerline{\epsfig{file=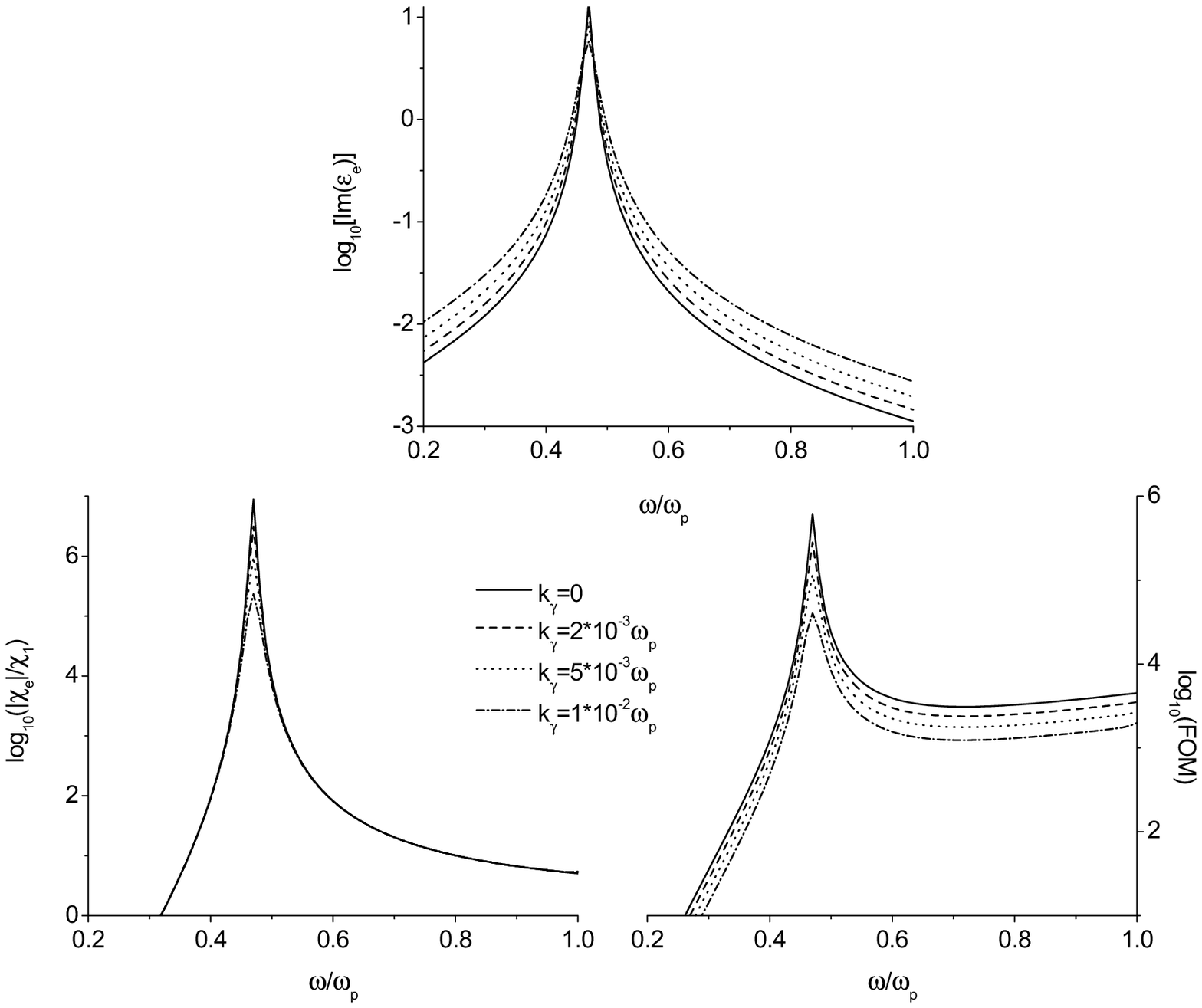,width=400pt}} \centerline{Fig.5.
/Gao, Huang and Yu}

\end{document}